# Multipole resonances and directional scattering by hyperbolic-media antennas

V.E. Babicheva[1]

[1]Center for Nano-Optics, Georgia State University, Atlanta, GA

baviev@gmail.com

**Abstract**. We propose to use optical antennas made out of natural hyperbolic material hexagonal boron nitride (hBN), and we demonstrate that this medium is a promising alternative to plasmonic and all-dielectric materials for realizing efficient subwavelength scatterers and metasurfaces based on them. We theoretically show that particles out of hyperbolic medium possess different resonances enabled by the support of high-k waves and their reflection from the particle boundaries. Among those resonances, there are electric quadrupole excitations, which cause magnetic resonance of the particle similar to what occurs in high-refractive-index particles. Excitations of the particle resonances are accompanied by the drop in the reflection from nanoparticle array to near-zero value, which can be ascribed to resonant Kerker effect. If particles are arranged in the spacer array with period d, narrow lattice resonances are possible at wavelength d, d/2, d/3 etc. This provides an additional degree of control and possibility to excite resonances at the wavelength defined by the array spacing. For the hBN particle with hyperbolic dispersion, we show that the full range of the resonances, including magnetic resonance and a decrease of reflection, is possible.

Optical antennas can be applied for efficient light control including enhancing light-matter interaction, increased localization, improving light harvesting, directional scattering, and so on [1]. In particular, plasmonic antennas have brought a great promise of light manipulation as they enable to localize light at subwavelength scale [2-4]. Various plasmonic nanostructures have been extensively studied for a wide range of functionality and applications [5-7]. Small particles of simple shapes, like spheres or disks, support only electric dipole and quadrupole resonances, and to achieve magnetic response one need to design nanostructures involving more complex shapes e.g. core-shell particles, split-ring resonators, or dielectric gaps between metal layers. Next, nanoantennas out of multilayer metal-dielectric structures have been proposed, and the excitation of both electric and magnetic resonances there have been studied [8-10]. Recently, all-dielectric antennas have emerged as a promising alternative to plasmonic nanostructures and, in particular, nanoparticles made out of high refractive index have attracted a lot of attention [11-15]. Particles of simple shapes like spheres or disks have both electric and magnetic resonances, and these resonances can spectrally overlap in the disks enabling unidirectional scattering [16].

Both plasmonic and all-dielectric structures have practical disadvantages and obstacles for real-life applications. Plasmonic structures are based on metal and suffer from high non-radiative losses. In all-dielectric structures such as silicon or germanium, non-radiative losses are low because of the small material losses, and radiative losses dominate preventing technology from miniaturization. Moreover, radiative losses can be especially high in the structures of a moderate refractive index such as oxides [17].

Here we propose to use antennas made out of natural hyperbolic material hexagonal boron nitride (hBN), and we demonstrate that this medium is a promising alternative to plasmonic and all-dielectric materials for realizing optical antennas and metasurfaces based on them. The hyperbolic medium has the different signs of real part of permittivity tensor components and consequently hyperbolic dispersion [18,19]. Artificially designed hyperbolic media (hyperbolic metamaterials or HMM) have been extensively studied in the last decade, and two main approaches for HMM realization have been widely used: arrays of metal nanorods and metal-dielectric multilayer structures.

Hyperbolic media support waves with high wavenumber (high-k waves), which enable a variety of interesting effects, like strong and broadband spontaneous emission enhancement, anomalous heat transfer, etc. [20-29]. Recently, high confinement of optical mode in the hyperbolic medium has been studied in waveguiding [30-32] and tapers [33]. HMM-based nanocavities and nanoresonators have been shown to support different modes [34-39], and to provide an increase of radiation efficiency, spasing, etc. [40-43]. However, because of the metal components of the structures, optical losses are high, resonances are broad, and performance of such plasmonic resonators is moderate.

Very recently, polar dielectrics, such as hBN [44-47], silicon carbide [48,49], gallium nitride [50], have been shown to support phonon-polariton resonances in the mid-infrared range, which goes with negative permittivity (so-called Reststrahlen band) and all effects similar to plasmonic ones but with much lower optical losses [51]. In polar dielectrics, polarization is induced by oscillating ions rather than electron-hole density. The lifetime of phonon-polaritons is much longer than plasmon-polaritons due to significantly larger scattering times associated with optical phonons as compared to plasmons. Furthermore, because of the van der Waals layered structure, hBN has resonances of in-plane and out-of-plane permittivity components at different wavelengths (two well-separated Reststrahlen bands, around 7 and 13 μm regions) and consequently possess hyperbolic dispersion with low optical losses. Thus, this natural hyperbolic material can be the main component of future photonic devices and applications [52-59].

Hexagonal boron nitride particles have been proposed as sub-wavelength resonators, and the possibility of multipole resonance excitations have been pointed out [60-62]. In the present work, for the first time, we study scattering properties of hBN particles, and in particular, we theoretically show that these particles support both electric and magnetic resonances, spectral overlaps of these resonances cause directional scattering (Kerker effect), and periodic arrangement of the particles can result in lattice resonances. The demonstrated effects provide the basis for further development of functional metasurfaces.

To start with, we consider antenna array with periods $d_x$ and $d_y$ and incident light polarized along the x-axis (Fig. 1a). Antenna dimensions are $a_x$, $a_y$, and $a_z$, and they are surrounded by the homogeneous environment with $\varepsilon_e = 1$. Antennas are made out of hyperbolic medium with permittivity components $\varepsilon_x = \varepsilon_y$, $Re(\varepsilon_x) < 0$ and $Re(\varepsilon_z) > 0$ (type II hyperbolic medium). First, we analyze properties of antennas that have $\varepsilon_x = \varepsilon_y = -14.6+1i$, and $\varepsilon_z = 2.7$, which corresponds to the hBN at wavelength 7 μm according to permittivity model of [63]. Throughout the study, we vary wavelength but keep permittivity value fix with the purpose to illustrate the general concept of resonance excitations. At the second part of the work, we show properties of the antennas with wavelength-dependent permittivity that corresponds to hBN.

For the considered hyperbolic medium, the propagation wavenumber ($k_x$, $k_y$, $k_z$) satisfies the dispersion equation

$$\frac{k_x^2}{\varepsilon_z} + \frac{k_y^2}{\varepsilon_z} + \frac{k_z^2}{\varepsilon_x} = k_0^2, \qquad (1)$$

where $k_0 = 2\pi/\lambda_0$ and $\lambda_0$ is free-space wavelength. One can expect that for the particle with dimensions $a_x$, $a_y$, and $a_z$, resonances occur when

$$k_x a_x = \pi n_x, \quad k_y a_y = \pi n_y, \quad k_z a_z = \pi n_z, \qquad (2)$$

where $n_x$, $n_y$, and $n_z$ are integer numbers.

Thus, in the resonance, the following condition is satisfied:

$$\frac{1}{\varepsilon_z}\left(\frac{n_x}{2a_x}\right)^2 + \frac{1}{\varepsilon_z}\left(\frac{n_y}{2a_y}\right)^2 + \frac{1}{\varepsilon_x}\left(\frac{n_z}{2a_z}\right)^2 = \frac{1}{\lambda_0^2}. \qquad (3)$$

In a more general case, Eq. (2) should include phase change at the boundary, e.g. $k_x a_x + 2\varphi_0 = \pi n_x$, but we restrict ourselves with the simplest model and show it works well.

Earlier, this model was shown to be applicable to describe resonances in the waveguides out of hyperbolic medium with finite-size ridges of HMM [30], nanowire resonators [34], and hyperbolic spheres [64]. This model of Fabry-Perot resonances has also been applied to dielectric nanobars [65]. In our case, because of the different signs of $Re(\varepsilon_x)$ and $Re(\varepsilon_z)$, both $k_x$ and $k_z$ wavevector components can be high (known as high-k waves in HMM), and the particle of hyperbolic medium with deeply subwavelength dimensions can support a number of different resonances.

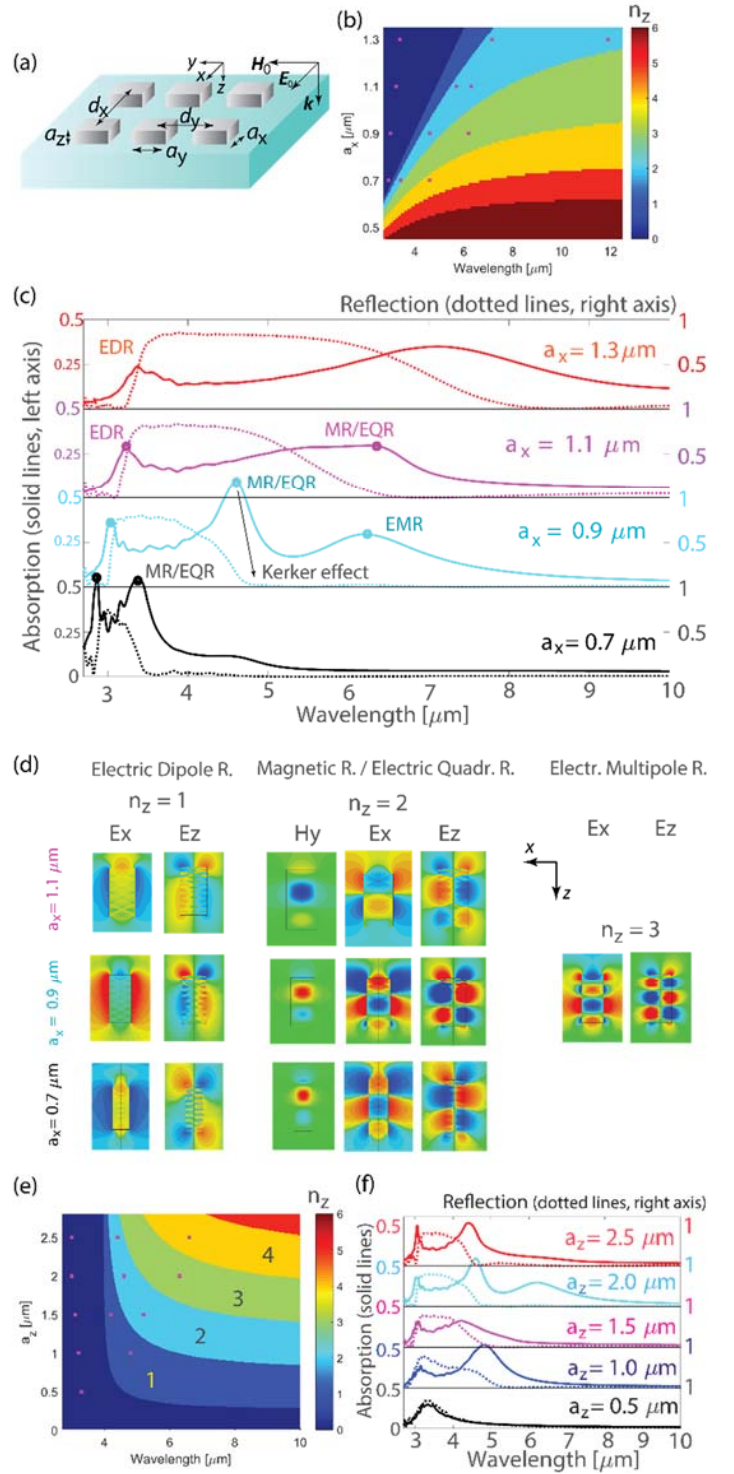

Fig. 1. Excitation of different resonances in the hyperbolic antennas. (a) Schematic of the structure. Antennas have dimensions $a_x$ x $a_y$ x $a_z$, arranged in the rectangular lattice with periods $d_x$ and $d_y$, and incident light polarized along the x-axis. (b) Colormap showing $n_z$ calculated by Eq. (3) for the different $a_x$. Antennas parameters are $a_y = 0.5$ μm, $a_z = 2$ μm, $n_y = 0$, and $n_x = 1$. Purple marks show the position of resonances in numerical modeling. (c) Numerical calculations of absorption and reflection spectra for the different $a_x$. Circle marks show peaks where field distribution is analyzed for electric dipole resonance (EDR), magnetic resonance (MR) and electric quadrupole resonances (EQR), and electric multipole resonance (EMR). Structure parameters are $d_x$ = 2.5 μm, $d_y$ = 2.5 μm, $a_y$ = 0.5 μm, $a_z$ = 2 μm, $\varepsilon_x = \varepsilon_y = -14.6+1i$, and $\varepsilon_z$ = 2.7. (d) Distributions of electric $E_x$, $E_z$, and magnetic $H_y$ fields at resonances. The magnetic field is shown only for $n_z$ = 2 where the magnetic resonance is present. Coordinate axes (XZ plane) correspond to all pictures. (e) Colormap showing $n_z$ calculated by Eq. (3) for different $a_z$. Antennas parameters are $a_x$ = 0.9 μm, $a_y$ = 0.5 μm, $n_y$ = 0, and $n_x$ = 1. Purple marks show the position of resonances in numerical modeling. (f) Numerical calculations of absorption and reflection spectra for different $a_z$. Structure parameters are $d_x$ = 2.5 μm, $d_y$ = 2.5 μm, $a_x$ = 0.9 μm, and $a_y$ = 0.5 μm.

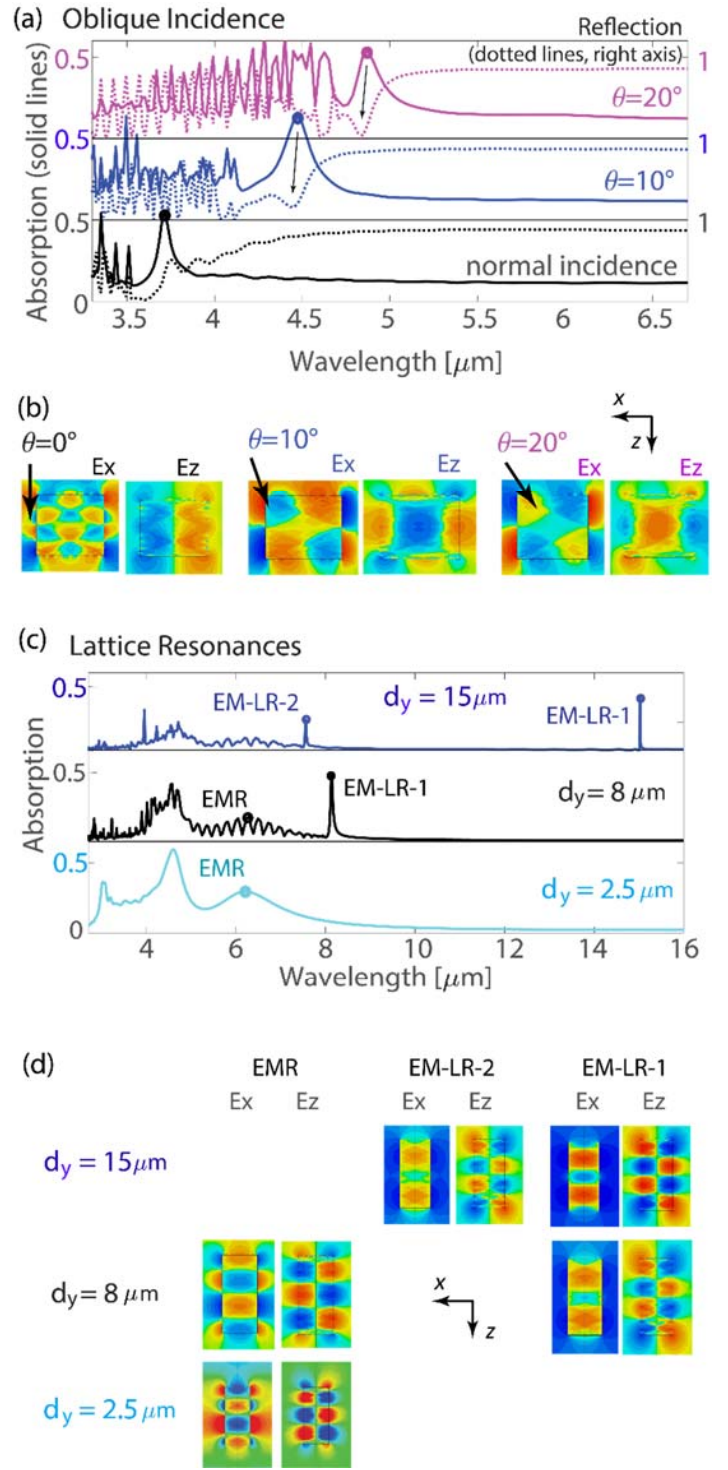

Comparison of results of Eq. (3) and numerical simulations with CST Microwave Studio, frequency domain solver shows a good agreement (Fig. 1). Excited resonances can be classified according to $n_z$: $n_z$ = 1 corresponds to electric dipole resonance (EDR); $n_z$ = 2 to electric quadrupole resonance (EQR); and we refer to the resonances with $n_z$ > 2 as electric multipole resonances (EMRs). Similar to the high-refractive-index particles, a reflection of the light from particle boundaries causes electric field circulation, and magnetic resonances (MR) are induced. The most efficiently MR is excited in the case of $n_z$ = 2, and we show mode field distribution for $H_y$ in those cases only (Fig. 1d). Calculations of $n_z$ with Eq. (3) provide a general guideline where electric and magnetic resonances can be expected for particular particle dimensions. Only EDRs fall out the model predictions and usually appear at the smaller wavelength. Higher multipoles are excited at a larger wavelength, which is different from plasmonic and dielectric particle resonances, where higher multipoles are at smaller wavelength range.

Excitation of MR/EQR is accompanied by the drop of reflection to nearly zero, which can be attributed to the destructive interference of light scattered backward by different multipoles and Kerker effect. Both MR and EQR are excited simultaneously and overlap with EDR, that is the generalized Kerker condition is satisfied [66-70].

Further, we consider larger particles with $a_x$ = 2.4 μm, to illustrate excitation of resonances with $n_x$>1 (Fig. 2). The resonance at λ = 3.7 μm at normal incidence is essentially different in comparison to the resonances at λ = 4.5 μm and oblique incidence with θ = 10° and λ = 4.9 μm at θ = 20°. The latter two correspond to $n_x$ = 2 and accompanied by the drop in reflection.

Lattice resonances in periodically arranged arrays are well studied for both plasmonic and all-dielectric particles [71-76]. In a similar way, particle out of hyperbolic medium can possess resonances at the wavelength defined by the period of the structure. Here, we show two examples: the array with $d_y$ = 8 μm has electric-multipole lattice resonance (EM-LR) excited at $λ_1$ ≈ $d_y$ = 8 μm, and the array with $d_y$ = 15 μm has two EM-LRs excited at $λ_1$ = $d_y$ = 15 μm and $λ_2$ = $d_y$/2 = 7.5 μm. From the field distribution, one can see that EM-LRs have higher $n_z$ then EMR (Fig. 2d). In other calculations (not shown here) we also observed resonances at $λ_3$ = $d_y$/3 and even $λ_4$ = $d_y$/4. A remarkable feature of lattice resonances is that they can be excited spectrally far away from the resonances of the single particle, and the example in Fig. 2c show well-pronounced resonances 8 μm further away from the last EMR in the closely spaced array.

Fig. 2. (a) Numerical calculations of absorption and reflection spectra for different angles of incidence. Circle marks show peaks where field distribution is analyzed. Structure parameters are $d_x$ = 3.5 μm, $d_y$ = 2.5 μm, $a_x$ = 2.4 μm, $a_y$ = 0.5 μm, and $a_z$ = 2 μm. (b) Distributions of $E_x$ and $E_z$ fields at resonances. Coordinate axes (XZ plane) correspond to all pictures. (c) Absorption spectra for the case of lattice resonances. For $d_y$ = 8 μm, electric-multipole lattice resonance (EM-LR) is excited at $λ_1$ ≈ $d_y$ = 8 μm, and for $d_y$ = 15 μm two EM-LRs are excited at $λ_1$ = $d_y$ = 15 μm and $λ_2$ = $d_y$/2 = 7.5 μm. Structure parameters are $d_x$ = 2.5 μm, $a_x$ = 0.9 μm, $a_y$ = 0.5 μm, and $a_z$ = 2 μm. (d) Distributions of $E_x$ and $E_z$ fields at resonances. Coordinate axes (XZ plane) correspond to all pictures. EMRs are excited with $n_z$ = 3, and EM-LRs with $n_z$ = 4.

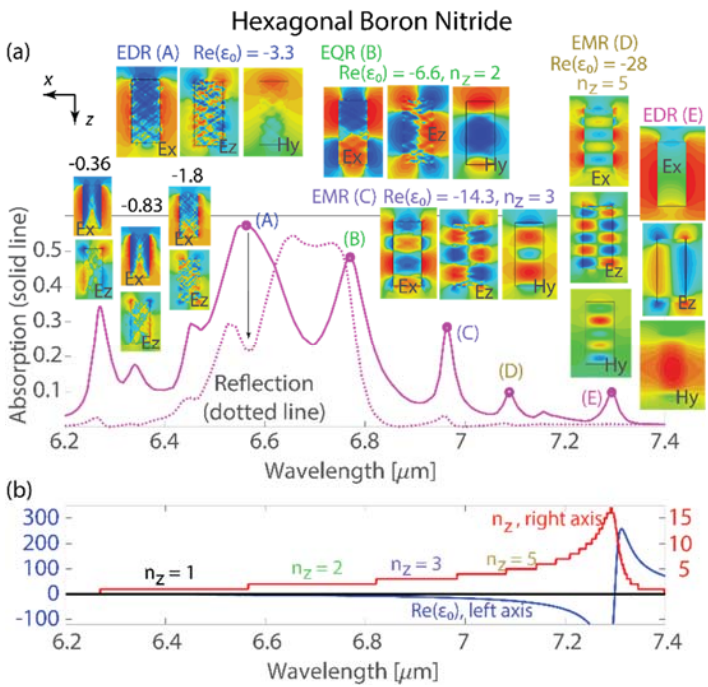

Fig. 3. Excitation of different resonances in hBN. (a) Absorption and reflection spectra as well as field distributions at the resonances. Structure parameters are $d_x$ = 2.5 µm, $d_y$ = 2.5 µm, $a_x$ = 0.9 µm, $a_y$ = 0.5 µm, and $a_z$ = 2 µm. (b) Real part of in-plane component of hBN permittivity (denoted $\varepsilon_0$ in figure) and $n_z$ calculated with Eq. (3). $n_y$ = 0 and $n_x$ = 1.

Finally, we analyze resonances and scattering properties of the antennas with realistic permittivity (Fig. 3; data and model from [63]). Within the band 6.2 – 7.4 µm, real part of in-plane permittivity component of hBN changes from near-zero to highly negative value followed by the sharp increase and positive value. In this case, a model of Eq. (3) also adequately predicts what resonances can be excited, and we observe multiple EDRs as well as full range of EMRs with $n_z$ = 2 to 5. Most of the resonances are accompanied by the near-zero reflection, but the reflection is the largest between EDR (A) and EQR (B).

To conclude, we have shown that particles out of hyperbolic medium possess different resonances enabled by the support of high-k waves and their reflection from the particle boundaries. These resonances are well predicted and described by the analytical expression Eq. (3). Among those resonances, there are electric quadrupole excitations, which cause magnetic resonance of the particle similar to what occurs in high-refractive-index particles. Excitations of the particle resonances are accompanied by the drop in the reflection from nanoparticle array to near-zero value, which can be ascribed to resonant Kerker effect. If particles are arranged in the spacer array with period $d_y$, narrow lattice resonances are possible at wavelength $d_y$, $d_y/2$, $d_y/3$ etc. This provides an additional degree of control and possibility to excite resonances at the wavelength defined by the array spacing. For the hBN particle with hyperbolic dispersion, we have shown that the full range of the resonances, including magnetic resonance and a decrease of reflection, is possible.

**References**

[1] Principles of Nano-Optics, L. Novotny, B. Hecht, Cambridge University Press, 2012

[2] E. Prodan, C. Radloff, N. J. Halas, P. Nordlander, "A Hybridization Model for the Plasmon Response of Complex Nanostructures," Science 302, 419-422 (2003).

[3] B. Luk'yanchuk, N.I. Zheludev, S. A. Maier, N. J. Halas, P. Nordlander, H. Giessen, and C. T. Chong, "The Fano resonance in plasmonic nanostructures and metamaterials," Nature Materials 9, 707–715 (2010).

[4] V.E. Babicheva, S.S. Vergeles, P.E. Vorobev, S. Burger, "Localized surface plasmon modes in a system of two interacting metallic cylinders," JOSA B 29, 1263-1269 (2012).

[5] H. A. Atwater, A. Polman, "Plasmonics for Improved Photovoltaic Devices," Nature Mater. 9, 205–213 (2010).

[6] A. Boulesbaa, V.E. Babicheva, K. Wang, I.I. Kravchenko, M.-W. Lin, M. Mahjouri-Samani, C. Jacob, A.A. Puretzky, K. Xiao, I. Ivanov, C.M. Rouleau, D.B. Geohegan, "Ultrafast Dynamics of Metal Plasmons Induced by 2D Semiconductor Excitons in Hybrid Nanostructure Arrays," ACS Photonics 3, 2389–2395 (2016).

[7] V.E. Babicheva, R.Sh. Ikhsanov, S.V. Zhukovsky, I.E. Protsenko, I.V. Smetanin, and A.V. Uskov, "Hot electron photoemission from plasmonic nanostructures: Role of surface photoelectric effect and transition absorption," ACS Photonics 2, 1039-1048 (2015).

[8] De Li, Ling Qin, Dong-Xiang Qi, Feng Gao, Ru-Wen Peng, Jin Zou, Qian-Jin Wang, and Mu Wang, "Tunable electric and magnetic resonances in multilayered metal/dielectric nanoplates at optical frequencies," J. Phys. D: Appl. Phys. 43, 345102 (2010).

[9] De Li, Ling Qin, Xiang Xiong, Ru-Wen Peng, Qing Hu, Guo-Bin Ma, Hao-Shen Zhou, and Mu Wang, "Exchange of electric and magnetic resonances in multilayered metal/dielectric nanoplates," Opt. Express 19, 22942-22949 (2011).

[10] Xiaoming Zhang, Jun-Jun Xiao, Qiang Zhang, Feifei Qin, Xingmin Cai, and Fan Ye, "Dual-Band Unidirectional Emission in a Multilayered Metal–Dielectric Nanoantenna," ACS Omega 2, 774–783 (2017).

[11] A. E. Krasnok, A. E. Miroshnichenko, P. A. Belov, Y. S. Kivshar, "All-Dielectric Optical Nanoantennas," Opt. Express 20, 20599-20604 (2012).

[12] A.I. Kuznetsov, A.E. Miroshnichenko, M.L. Brongersma, Y.S. Kivshar, B. Luk'yanchuk, "Optically resonant dielectric nanostructures," Science 354, aag2472 (2016).

[13] I. Staude and J. Schilling, "Metamaterial-inspired silicon nanophotonics," Nature Photonics 11, 274–284 (2017).

[14] K.V. Baryshnikova, M.I. Petrov, V.E. Babicheva, P.A. Belov, "Plasmonic and silicon spherical nanoparticle antireflective coatings," Scientific Reports 6, 22136 (2016).

[15] V. Babicheva, M. Petrov, K. Baryshnikova, P. Belov, "Reflection compensation mediated by electric and magnetic resonances of all-dielectric metasurfaces [Invited]," Journal of the Optical Society of America B 34 (7), D18-D28 (2017).

[16] I. Staude, A. E. Miroshnichenko, M. Decker, N. T. Fofang, S. Liu, E. Gonzales, J. Dominguez, T. S. Luk, D. N. Neshev, I. Brener, Y. Kivshar, "Tailoring Directional Scattering through Magnetic and Electric Resonances in Subwavelength Silicon Nanodisks," ACS Nano 7, 7824–7832 (2013).

[17] S. Zhang, R. Jiang, Y. M. Xie, Q. Ruan, B. Yang, J. Wang, H. Q. Lin, "Colloidal Moderate-Refractive-Index Cu2O Nanospheres as Visible-Region Nanoantennas with Electromagnetic Resonance and Directional Light-Scattering Properties," Adv. Mater. 27(45), 7432-7439 (2015).

[18] V. P. Drachev, V. A. Podolskiy, and A. V. Kildishev, "Hyperbolic metamaterials: new physics behind a classical problem," Opt. Express 21(12), 15048–15064 (2013).

[19] A. Poddubny, I. Iorsh, P. Belov, and Y. Kivshar, "Hyperbolic metamaterials," Nat. Photonics 7(12), 948–957 (2013).

[20] H. N. S. Krishnamoorthy, Z. Jacob, E. Narimanov, I. Kretzschmar, and V. M. Menon, "Topological transitions in metamaterials," Science 336(6078), 205–209 (2012).

[21] S. V. Zhukovsky, O. Kidwai, and J. E. Sipe, "Physical nature of volume plasmon polaritons in hyperbolic metamaterials," Opt. Express 21(12), 14982–14987 (2013).

[22] Z. Jacob, I. I. Smolyaninov, and E. E. Narimanov, "Broadband Purcell effect: Radiative decay engineering with metamaterials," Appl. Phys. Lett. 100, 181105 (2012).

[23] Z. Jacob, J.-Y. Kim, G. V. Naik, A. Boltasseva, E. E. Narimanov, and V. M. Shalaev, "Engineering photonic density of states using metamaterials," Appl. Phys. B 100, 215 (2010).

[24] Z. Liu, H. Lee, Y. Xiong, C. Sun, and X. Zhang, "Far-field optical hyperlens magnifying sub-diffraction-limited objects," Science 315(5819), 1686 (2007).

[25] S.V. Zhukovsky, A.A. Orlov, V.E. Babicheva, A.V. Lavrinenko, J.E. Sipe, "Photonic-band-gap engineering for volume plasmon polaritons in multiscale multilayer hyperbolic metamaterials," Physical Review A 90, 013801 (2014).

[26] C. Simovski, S. Maslovski, I. Nefedov, S. Tretyakov, "Optimization of radiative heat transfer in hyperbolic metamaterials for thermophotovoltaic applications," Opt. Express 21, 14988 (2013).

[27] A.A. Orlov, A.K. Krylova, S.V. Zhukovsky, V.E. Babicheva, P.A. Belov, "Multi-periodicity in plasmonic multilayers: general description and diversity of topologies," Phys. Rev. A 90, 013812 (2014).

[28] A.A. Orlov, E.A. Yankovskaya, S.V. Zhukovsky, V.E. Babicheva, I.V. Iorsh, P.A. Belov, "Retrieval of Effective Parameters of Subwavelength Periodic Photonic Structures," Crystals 4, 417-426 (2014).

[29] A.V. Chebykin, V.E. Babicheva, I.V. Iorsh, A.A. Orlov, P.A. Belov, S.V. Zhukovsky, "Enhancement of the Purcell factor in multiperiodic hyperboliclike metamaterials," Physical Review A 93, 033855 (2016).

[30] V. E. Babicheva, M. Y. Shalaginov, S. Ishii, A. Boltasseva, and A. V. Kildishev, "Finite-width plasmonic waveguides with hyperbolic multilayer cladding," Opt. Express 23(8), 9681–9689 (2015).

[31] S. Ishii, M. Y. Shalaginov, V. E. Babicheva, A. Boltasseva, and A. V. Kildishev, "Plasmonic waveguides cladded by hyperbolic metamaterials," Opt. Lett. 39(16), 4663–4666 (2014).

[32] V.E. Babicheva, M.Y. Shalaginov, S. Ishii, A. Boltasseva, and A.V. Kildishev, "Long-range plasmonic waveguides with hyperbolic cladding," Opt. Express 23, 31109-31119 (2015).

[33] P.R. West, N. Kinsey, M. Ferrera, A.V. Kildishev, V.M. Shalaev, and A. Boltasseva, "Adiabatically Tapered Hyperbolic Metamaterials for Dispersion Control of High-k Waves," Nano Lett. 15(1), 498–505 (2015).

[34] Jie Yao, Xiaodong Yang, Xiaobo Yin, Guy Bartal, and Xiang Zhang, "Three-dimensional nanometer-scale optical cavities of indefinite medium," PNAS 108, 11327–11331 (2011).

[35] Xiaodong Yang, Jie Yao, Junsuk Rho, Xiaobo Yin, and Xiang Zhang, "Experimental realization of three-dimensional indefinite cavities at the nanoscale with anomalous scaling laws" Nature Photonics 6, 450–454 (2012).

[36] Chihhui Wu, Alessandro Salandrino, Xingjie Ni, and Xiang Zhang, "Electrodynamical Light Trapping Using Whispering-Gallery Resonances in Hyperbolic Cavities," Phys. Rev. X 4, 021015 (2014).

[37] Yingran He, Huixu Deng, Xiangyang Jiao, Sailing He, Jie Gao, and Xiaodong Yang, "Infrared perfect absorber based on nanowire metamaterial cavities," Opt. Lett. 38, 1179-1181 (2013)

[38] S. Ishii, S. Inoue, and A. Otomo, "Scattering and absorption from strongly anisotropic nanoparticles," Opt. Express 21, 23181-23187 (2013).

[39] R. Smaali, F. Omeis, A. Moreau, E. Centeno, and T. Taliercio "Miniaturizing optical antennas using hyperbolic metamaterial wires," Phys. Rev. B 95, 155306 (2017).

[40] C. Guclu, T. S. Luk, G. T. Wang, and F. Capolino, "Radiative emission enhancement using nano-antennas made of hyperbolic metamaterial resonators," Applied Physics Letters 105, 123101 (2014).

[41] M. Hasan and I. Iorsh, "Interaction of light with a hyperbolic cavity in the strong-coupling regime with Fano resonance," 2015 Days on Diffraction (DD), St. Petersburg, pp. 1-6, doi: 10.1109/DD.2015.7354848 (2015).

[42] A.P. Slobozhanyuk, P. Ginzburg, D.A. Powell, I. Iorsh, A.S. Shalin, P. Segovia, A.V. Krasavin, G.A. Wurtz, V.A. Podolskiy, P.A. Belov, and A.V. Zayats, "Purcell effect in hyperbolic metamaterial resonators," Phys. Rev. B 92, 195127 (2015).

[43] Mingjie Wan, Ping Gu, Weiyue Liu, Zhuo Chen, and Zhenlin Wang, "Low threshold spaser based on deep-subwavelength spherical hyperbolic metamaterial cavities," Appl. Phys. Lett. 110, 031103 (2017).

[44] J. D. Caldwell, I. Vurgaftman, J. G. Tischler, O. J. Glembocki, J. C. Owrutsky, and T. L. Reinecke, "Atomic-scale photonic hybrids for mid-infrared and terahertz nanophotonics," Nature Nanotechnology 11, 9–15 (2016).

[45] S. Dai, Z. Fei, Q. Ma, A.S. Rodin, M. Wagner, A.S. McLeod, M.K. Liu, W. Gannett, W. Regan, K. Watanabe, T. Taniguchi, M. Thiemens, G. Dominguez, A.H. Castro Neto, A. Zettl, F. Keilmann, P. Jarillo-Herrero, M. M. Fogler, D. N. Basov, "Tunable Phonon Polaritons in Atomically Thin van der Waals Crystals of Boron Nitride," Science 343, 1125 (2014).

[46] P. Li, I. Dolado, F. J. Alfaro-Mozaz, A. Yu. Nikitin, F. Casanova, L. E. Hueso, S. Vélez, R. Hillenbrand, "Optical Nanoimaging of Hyperbolic Surface Polaritons at the Edges of van der Waals Materials," Nano Lett. 17, 228–235 (2017).

[47] S. Dai, Q. Ma, M. K. Liu, T. Andersen, Z. Fei, M. D. Goldflam, M. Wagner, K. Watanabe, T. Taniguchi, M. Thiemens, F. Keilmann, G. C. A. M. Janssen, S-E. Zhu, P. Jarillo-Herrero, M. M. Fogler, D. N. Basov, "Graphene on hexagonal boron nitride as a tunable hyperbolic metamaterial," Nature Nanotechnology 10, 682–686 (2015).

[48] J.D. Caldwell, O.J. Glembocki, Y. Francescato, N. Sharac, V. Giannini, F. J. Bezares, J. P. Long, J. C. Owrutsky, I. Vurgaftman, J. G. Tischler, V. D. Wheeler, N. D. Bassim, L. M. Shirey, R. Kasica, and S. A. Maier, "Low-Loss, Extreme Subdiffraction Photon Confinement via Silicon Carbide Localized Surface Phonon Polariton Resonators," Nano Lett., 2013, 13 (8), pp 3690–3697

[49] C.T. Ellis, J.G. Tischler, O.J. Glembocki, F.J. Bezares, A.J. Giles, R. Kasica, L. Shirey, J.C. Owrutsky, D.N. Chigrin, and J.D. Caldwell, "Aspect-ratio driven evolution of high-order resonant modes and near-field distributions in localized surface phonon polariton nanostructures," Scientific Reports 6, 32959 (2016).

[50] K. Feng, W. Streyer, S. M. Islam, J. Verma, D. Jena, D. Wasserman, and A.J. Hoffman, "Localized surface phonon polariton resonances in polar gallium nitride," Appl. Phys. Lett. 107, 081108 (2015).

[51] J.D. Caldwell, L. Lindsay, V. Giannini, I. Vurgaftman, T.L. Reinecke, S.A. Maier, and O.J. Glembocki, "Low-loss, infrared and terahertz

nanophotonics using surface phonon polaritons," Nanophotonics 4, 44–68 (2015).

[52] J. Sun, N. M. Litchinitser, and J. Zhou, "Indefinite by Nature: From Ultraviolet to Terahertz," ACS Photonics 1(4), 293–303 (2014).

[53] E. E. Narimanov and A. V. Kildishev, "Naturally hyperbolic," Nat. Photonics 9(4), 214–216 (2015).

[54] Bo Zhao and Zhuomin M. Zhang, "Resonance perfect absorption by exciting hyperbolic phonon polaritons in 1D hBN gratings," Opt. Express 25, 7791-7796 (2017).

[55] A.Yu. Nikitin, E. Yoxall, M. Schnell, S. Vélez, I. Dolado, P. Alonso-Gonzalez, F. Casanova, L. E. Hueso, and R. Hillenbrand, "Nanofocusing of Hyperbolic Phonon Polaritons in a Tapered Boron Nitride Slab," ACS Photonics 3(6), 924–929 (2016).

[56] P. Li, M. Lewin, A. V. Kretinin, J. D. Caldwell, K. S. Novoselov, T. Taniguchi, K. Watanabe, F. Gaussmann, and T. Taubner, "Hyperbolic phonon-polaritons in boron nitride for near-field optical imaging and focusing," Nature Commun. 6, 7507 (2015).

[57] S. Dai, Q. Ma, T. Andersen, A.S. Mcleod, Z. Fei, M.K. Liu, M. Wagner, K. Watanabe, T. Taniguchi, M. Thiemens, F. Keilmann, P. Jarillo-Herrero, M.M. Fogler, D.N. Basov, "Subdiffractional focusing and guiding of polaritonic rays in a natural hyperbolic material," Nature Commun. 6, 6963 (2015).

[58] P. Li, I. Dolado, F. J. Alfaro-Mozaz, A. Yu. Nikitin, F. Casanova, L. E. Hueso, S. Vélez, and R. Hillenbrand, "Optical Nanoimaging of Hyperbolic Surface Polaritons at the Edges of van der Waals Materials," Nano Lett. 17(1), 228–235 (2017).

[59] F. J. Alfaro-Mozaz, P. Alonso-González, S. Vélez, I. Dolado, M. Autore, S. Mastel, F. Casanova, L. E. Hueso, P. Li, A. Y. Nikitin, and R. Hillenbrand, "Nanoimaging of resonating hyperbolic polaritons in linear boron nitride antennas," Nature Communications 8, 15624 (2017).

[60] J.D. Caldwell, A.V. Kretinin, Y. Chen, V. Giannini, M.M. Fogler, Y. Francescato, C.T. Ellis, J.G. Tischler, C.R. Woods, A.J. Giles, M. Hong, K. Watanabe, T. Taniguchi, S.A. Maier, and K.S. Novoselov, "Sub-diffractional volume-confined polaritons in the natural hyperbolic material hexagonal boron nitride," Nature Commun. 5, 5221 (2014).

[61] A.J. Giles, S. Dai, O.J. Glembocki, A.V. Kretinin, Z. Sun, C.T. Ellis, J.G. Tischler, T. Taniguchi, K. Watanabe, M. M. Fogler, K.S. Novoselov, D. N. Basov, and J.D. Caldwell, "Imaging of Anomalous Internal Reflections of Hyperbolic Phonon-Polaritons in Hexagonal Boron Nitride," Nano Lett. 16(6), 3858–3865 (2016).

[62] Z. Sun, Á. Gutierrez-Rubio, D. N. Basov, and M. M. Fogler, "Hamiltonian Optics of Hyperbolic Polaritons in Nanogranules," Nano Lett. 15, 4455–4460 (2015).

[63] Y. Cai, L. Zhang, Q. Zeng, L. Cheng, Y. Xu, "Infrared reflectance spectrum of BN calculated from first principles," Solid State Communications 141, 262–266 (2007).

[64] S. Ishii, S. Inoue, and A. Otomo, "Electric and magnetic resonances in strongly anisotropic particles," J. Opt. Soc. Am. B 31, 218-222 (2014).

[65] Y. Yang, A. E. Miroshnichenko, S.V. Kostinski, M. Odit, P.Kapitanova, M. Qiu, and Y.S. Kivshar, "Multimode directionality in all-dielectric metasurfaces," Phys. Rev. B 95, 165426 (2017).

[66] M. Kerker, D. Wang, C. Giles, "Electromagnetic Scattering by Magnetic Spheres," J. Opt. Soc. Am. 73, 765 (1983).

[67] Y. H. Fu, A. I. Kuznetsov, A. E. Miroshnichenko, Y. F. Yu, B. Luk'yanchuk, "Directional Visible Light Scattering by Silicon Nanoparticles," Nat. Commun. 4, 1527 (2013).

[68] S. Person, M. Jain, Z. Lapin, J. J. Sáenz, G. Wicks, L. Novotny, "Demonstration of Zero Optical Backscattering from Single Nanoparticles," Nano Lett. 13(4), 1806–1809 (2013).

[69] A. Pors, S. K. H. Andersen, and S. I. Bozhevolnyi, "Unidirectional scattering by nanoparticles near substrates: generalized Kerker conditions," Opt. Express 23, 28808-28828 (2015).

[70] R. Alaee, R. Filter, D. Lehr, F. Lederer, and C. Rockstuhl, "A generalized Kerker condition for highly directive nanoantennas," Opt. Lett. 40, 2645-2648 (2015).

[71] B. Auguié and W.L. Barnes "Collective resonances in gold nanoparticle arrays," Phys. Rev. Lett. 10, 143902 (2008).

[72] A.B. Evlyukhin, C. Reinhardt, U. Zywietz, B. Chichkov, "Collective resonances in metal nanoparticle arrays with dipole-quadrupole interactions," Phys. Rev. B 85(24), 245411 (2012).

[73] S. V. Zhukovsky, V. E. Babicheva, A. V. Uskov, I. E. Protsenko, and A. V. Lavrinenko, E"nhanced Electron Photoemission by Collective Lattice Resonances in Plasmonic Nanoparticle-Array Photodetectors and Solar Cells," Plasmonics 9, 283 (2014).

[74] S. V. Zhukovsky, V. E. Babicheva, A. V. Uskov, I. E. Protsenko, and A. V. Lavrinenko, "Electron Photoemission in Plasmonic Nanoparticle Arrays: Analysis of Collective Resonances and Embedding Effects," Appl. Phys. A 116, 929 (2014).

[75] A. B. Evlyukhin, C. Reinhardt, A. Seidel, B. S. Luk'yanchuk,; B. N. Chichkov, "Optical Response Features of Si-Nanoparticle Arrays," Phys. Rev. B 82(4), 045404 (2010).

[76] A.B. Evlyukhin and V.E. Babicheva, "Resonant lattice Kerker effect in metasurfaces with electric and magnetic optical responses", https://arxiv.org/abs/1705.05533, 2017.